\documentclass[a4paper,12pt]{revtex4}
\usepackage[dvips]{graphicx}
\usepackage{delarray}
\usepackage{amsmath}
\usepackage{amssymb}
\usepackage{here}
\newcommand{\mygtrsim}{\mathrel{\mbox{\raisebox{-1mm}{$\stackrel{>}{\sim}$}}}}

\graphicspath{{pictures/}}
\usepackage{ae}
\usepackage{color}
\usepackage{ifthen}
\usepackage{bm} \newcommand{\g}[1]{\mbox{\boldmath $#1$}}
\usepackage{ifpdf}
\ifpdf
	\usepackage[pdftex]{hyperref}
	\hypersetup{colorlinks=true,
		pdfstartview=FitV,
		linkcolor=blue,
		citecolor=red,
		urlcolor=blue,
	}
	\DeclareGraphicsExtensions{.pdf}
\else
	\usepackage[dvips]{graphicx}
	\usepackage{epsfig}
	\DeclareGraphicsExtensions{.eps}
	\usepackage{url}
\fi

\newcommand{\beq}{\begin{equation}}
\newcommand{\eeq}{\end{equation}}

\begin{document}

\begin{center}
\Large

{\bf Alfv\'enic instabilities driven by runaways in fusion plasmas} \\
~\\*[-0.5cm] \normalsize {  T.~F\"ul\"op$^1$ and S.~Newton$^2$
  \\\it\small
  $^1$ Department of Applied Physics, Chalmers University of
  Technology, G\"oteborg, Sweden\\ $^2$ Euratom/CCFE Fusion Association, Culham Science Centre,
  Abingdon OX14 3DB, UK}\\ \today
\end{center}
\begin{abstract}
  \vspace{-0.5cm} Runaway particles can be produced in plasmas with
  large electric fields.  Here we address the possibility that such runaway
 ions and electrons excite Alfv\'enic instabilities.  The magnetic
  perturbation induced by these modes can enhance the loss of
  runaways.  This may have important implications for the runaway
  electron beam formation in tokamak disruptions.
\end{abstract}
\maketitle {\em Introduction} Runaway electron (RE) generation in the
presence of electric fields is common in both laboratory and space
plasmas \cite{re}.  This can occur because the friction force
experienced by an electron due to Coulomb collisions is a
non-monotonic function, having a maximum at the thermal speed and
decreasing at higher speeds.  For sufficiently fast electrons, an
accelerating electric force can overcome this friction and the
electrons can then run away.  In laboratory plasmas, much attention
has been given to the highly relativistic RE beams that can be
generated in tokamak disruptions.  Such REs may damage plasma facing
components due to their highly localized energy deposition.  The
potential for detrimental effects increases with plasma
current. Therefore, understanding the processes that may eliminate RE
beam formation is very important for future reactor-scale tokamaks
with high currents, such as ITER \cite{lehnen}.  In several tokamak
experiments it has been observed that RE generation only occurs above
a threshold toroidal magnetic field \cite{gill,jt60}.  While the
origin of this threshold is uncertain, it has been linked to decreased
relative magnetic fluctuation levels \cite{jt60,kudyakov}.  Recent
work {at}
the TEXTOR tokamak \cite{zeng} has shown the presence of
fluctuations in the frequency range $f \simeq 60-260\;\rm kHz$, {during}
disruptions deliberately triggered by the injection of argon.  The
presence of these fluctuations appears to be instrumental in limiting
the RE beam formation in these cases. The aim of this work is to
investigate Alfv\'enic instabilities driven by suprathermal ions and
electrons in suddenly cooling impure plasmas and their possible
connection with the fluctuations observed in tokamak disruptions.

{There are many observations}  that Alfv\'en waves can be driven
 unstable via particle resonance both in natural and laboratory
 plasmas \cite{heidbrink,hughes,zonca}.  At low frequency, the free
 energy originates from an inverted energy distribution or spatial
 inhomogeneity.  The resonance condition requires that particles
 achieve a significant, well defined fraction of the Alfv\'{e}n
 velocity $v_A=B/\sqrt{\mu_0 \rho_m}$, where $\rho_m$ is the mass
 density.  As we will show here, this becomes increasingly difficult
 as the magnetic field $B$ increases, and could provide a possible
 explanation for the experimentally observed threshold in the magnetic
 perturbation level described in \cite{jt60,zeng,kudyakov}. Runaway
 ions are expected to have inverted energy distributions, which will
 drive Alfv\'enic instabilities if they attain sufficient velocity to
 fulfil the resonance condition.  Whilst runaway electron energy
 distributions are rarely inverted, Alfv\'enic instabilities can also
 be driven by resonant interaction with fast electrons with {steep}
 density profiles.  Whichever the drive, the appearance of such
 instabilities in tokamak disruptions can have important consequences.
 The magnetic perturbations associated with the wave can scatter the
 runaway electrons and terminate the beam~{\cite{peretal2000}},
 providing a passive mitigation of the detrimental effects of the RE
 beams.  Alfv\'enic instabilities can also be used as a diagnostic for
 the plasma, through the technique of MHD spectroscopy \cite{holties}.

In tokamaks, one of the most important Alfv\'enic instabilities is the
Toroidal Alfv\'en Eigenmode (TAE) \cite{chengchance}. TAEs are
discrete modes residing in toroidicity induced gaps in the shear
Alfv\'{e}n continuum, and are therefore usually only weakly damped, as
they are not subject to continuum damping.  Interestingly, TAEs can
have frequencies and mode numbers in the same range as the
experimental observations in Ref.~\cite{zeng}. TAE modes have been
shown to be driven unstable by a wide variety of energetic ion
populations, including fast ions produced by neutral beam injection or
ion cyclotron resonance heating and alpha-particles produced in DT
fusion reactions \cite{heidbrink}.  Here we determine the distribution
function of high energy ions generated by the large electric field
that accelerates the runaway electrons.  We then consider the conditions
under which the runaway ions and electrons can produce TAE growth and
discuss the connection with the experimental observations.

{\it Ion runaway}
Runaway acceleration of ions in the presence of an electric field has
been considered during magnetic reconnection events in tokamaks
\cite{helanderion}, solar flares \cite{holman} and lightning
discharges \cite{fulop}.  To allow ion runaway, the frictional drag
due to the drifting electrons should not cancel the electric
force. This is the case in the presence of magnetic trapping
or impurities with a different charge to that of the ions \cite{fr}.
{Bulk ions interacting via Coulomb collisions in a plasma
  experience a non-monotonic friction force. Drag against ions
  dominates at low energy, decreases with velocity to a minimum at
  $v_m$ and increases at energies exceeding this as drag against
  electrons takes over.}  If the ion speed is much lower than the
thermal electron speed, $v \ll v_{Te}$, in a straight magnetic field
the condition for ions (i) to be accelerated when moving in a
Maxwellian distribution of field particles (j) is \cite{fulop}
$
{E}/{E_D}>\left[\sum_j( n_jZ_iZ^2_j T_e)/(n_eT_j)\left(1+m_j/m_i\right)G\left(v/v_{Tj}\right)\right]/{|1-Z_i/Z_{\rm eff} |}.
$
Throughout $m_s$, $T_s$, $n_s$ and $Z_s e$ denote respectively the
mass, temperature, number density and charge of particles of species
$s$, $v_{Te} = \sqrt{2T_e/m_e}$, $E_D=(n_ee^3 \ln{\Lambda})/(4
\pi\epsilon_0^2 T_e)$ is the Dreicer field, $\ln\Lambda$ is the
Coulomb logarithm, $Z_{\rm eff}=n_e^{-1}\sum_jn_jZ_j^2$ is the
effective charge (where the summation is over all ion species) and $G(x)$
is the Chandrasekhar function. Neutrals are not expected to
penetrate the runaway electron beam \cite{hollmann}, so in this work
friction due to collisions with neutral particles will be neglected.
\begin{figure}[htbp]
\begin{center}
  \includegraphics[width=0.49\textwidth]{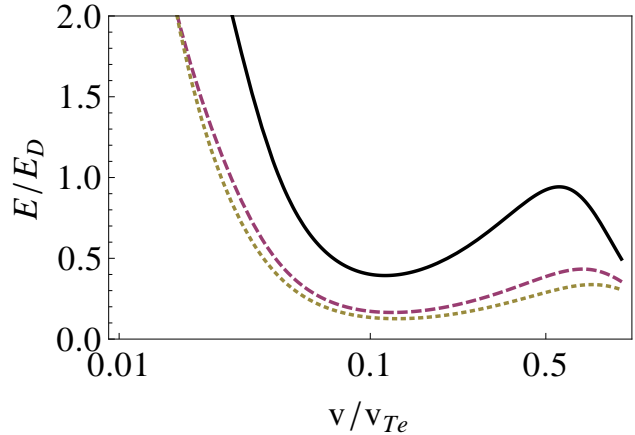} 
\caption{ 
    Minimum accelerating field $E/E_D$ as a function of the normalized
    deuterium ion speed for varying impurity content. Solid: $n_{Ar}=
    0.1 n_D$, $n_C=0.02 n_D$, $Z_{Ar}=Z_C=2$, $Z_{\rm
      eff}=1.2$. Dashed: $n_{Ar}= 0.1 n_D$, $n_C=0.08 n_D$,
    $Z_{Ar}=Z_C=3$, $Z_{\rm eff}=1.7$. Dotted $n_{Ar}= 0.5 n_D$,
    $n_C=0.08n_D$, $Z_{Ar}=Z_C=3$, $Z_{\rm eff}=2.3$.} 
       \label{fig_e}
\end{center}
\end{figure}
The minimum acceleration field $E/E_D$ is shown as a function of
normalized ion speed in Fig.~\ref{fig_e}, for deuterium ions in
scenarios similar to that described in \cite{zeng}, where a disruption
was triggered by the injection of argon particles.  The post
disruption temperature is often poorly diagnosed and we have taken a
characteristic value of 10 eV for all species, as at such a low
temperature the species may be expected to equilibrate.  At these
temperatures the impurities are not fully ionized, and the charge
states vary rapidly with time and space. During the disruption itself,
the impurities mix into the core and reach high charge states during
the cooling phase. They recombine during the current quench but with a
possible slight time delay. Here, the heavy argon impurities are taken
to have charge $Z_{Ar}=2$ or $Z_{Ar}=3$, and a typical background
carbon impurity with the same charge was also assumed.  There is a
clear minimum in the collisional friction on the deuterium ions at
around 10\% of the electron thermal speed, which is robust to the
variation in impurity content.  If the electric field is large enough,
deuterium ions from the tail of the thermal ion distribution will be
accelerated.  The instability growth rate will depend on the details
of the resulting fast ion distribution, which can be found by solving
the kinetic equation for high energy ions.
 
To allow the study of time-dependent situations, such as a disruptive
instability where a steady state is not likely to be established, the
kinetic equation must be solved as an initial-value problem.  Such a
solution was outlined in Ref.~\cite{helanderion} in the limit of trace
impurities.  Here we generalise this calculation for arbitrary
impurity content.  Assuming the friction is dominated by Coulomb
collisions, the kinetic equation for the ion distribution function $f$
can be written as
${\partial f}/{\partial t}+v_\parallel \nabla_\parallel f+\g
v_d\cdot \nabla f+({Z_ieE_\ast}/{m_i})({v_\parallel}/{v}) ({\partial f}/{\partial
  v})=C(f), $ 
where the subscript $\parallel$ is taken with respect to the
background magnetic field and ${\bf v}_d$ is the magnetic drift due to
the field inhomogeneity.  The effective electric force
$eE_\ast=eE_\parallel + R_{ie\parallel}/n_iZ_i$ is the sum of the
electric force and the average friction force on the ions from the
electrons.  At the low temperatures of interest, we neglect trapping
effects and it can be shown by invoking momentum conservation that
$eE_\parallel=-Z_{\rm eff}R_{ie\parallel}/(n_iZ_i^2)$, so that
$E_\ast=E_\parallel(1-Z_i/Z_{\rm eff})$.  Note that in a pure plasma,
when $Z_i=Z_{\rm eff}$, the effective field is zero and the test ion
will always slow down.  The operator $C(f)$ represents collisions
between the fast ions and thermal background Maxwellian ions and
electrons:
\begin{eqnarray}
  C(f)=\frac{Z_{\rm eff}v_c^3}{2 v^3 \tau_s}\frac{\partial}{\partial \xi}(1-\xi^2)\frac{\partial f }{\partial \xi}+\frac{1}{v^2\tau_s}\frac{\partial }{\partial v}\left[ \left(\bar{n}v_c^3+v^3\right)f+\left(\bar{n}v_c^3 +\frac{T_e}{T_i}v^3\right)\frac{T_i}{m_i v}\frac{\partial f}{\partial v}\right].
\end{eqnarray}
Here $\xi=v_\parallel/v$ is the pitch, the critical velocity for ion
slowing on electrons is $ v_c=\left(3 \sqrt{\pi}m_e/4 m_i\right)^{1/3}
v_{Te}$, the characteristic time for fast ion slowing on electons
$\tau_s$ is written in terms of the ion self-collision time $\tau_{ii}
= 3 (2 \pi)^{3/2}\epsilon_0^2 \sqrt{m_i} T_i^{3/2} / n_i Z_i^4 e^4
\ln{\Lambda}$,
$$
\tau_s=\left(\frac{m_i}{m_e}\right)^{1/2}\left(\frac{T_e}{T_i}\right)^{3/2}\frac{n_iZ_i^2}{n_e}\tau_{ii}, \qquad
\bar{n}=\frac{n_iZ_i^2}{n_e}\left(1+\sum_{j\neq i}\frac{n_jZ_j^2m_i}{n_iZ_i^2m_j}\right).
$$ In the trace impurity limit $\bar{n}=Z_i$.
{For the cold plasmas of interest here $Z_z m_i / m_z < 1$, so
  $\bar{n}$ is always less than one if the main ions are hydrogenic.}
{For illustration, we note that minimising the dynamic
  friction~\cite{helanderion}, which is} approximately proportional to
$\bar{n}v_c^3/v^2+v$ gives $v_m=(2 \bar{n})^{1/3}v_c$, with
corresponding kinetic energy $E_m=m_iv_m^2/2=\left(9 \pi m_i/4
  m_e\right)^{1/3} \bar{n}^{2/3}T_e.$ For deuterium ions $E_m \simeq
30 \bar{n}^{2/3}T_e$ and thus they can be accelerated to high energies
unless $\bar{n}$ is very low.

In the limit of $\delta=E_\ast T_i/E_D T_e\ll 1$ the solution of the
kinetic equation can be obtained by an expansion $
F=\ln{f}=\delta^{-1}F^{(0)}+\delta^{-1/2}F^{(1)}+\ldots $. Rescaling
the independent variables by writing $\tau=3
\delta^{3/2}(\pi/2)^{1/2}(n_e/n_iZ_i^2\tau_{ii})t$ and $w=v(\delta
m_i/T_i)^{1/2}$, the runaway ion distribution function may be given
approximately as~\cite{helanderion}
\begin{equation}
f_{RI}(w,\xi, \tau) \propto\exp{\left[-\frac{w^2}{2\delta}+\frac{w^4-(w^3-3
   \bar{n}\tau)^{4/3}H(w^3-3\bar{n}\tau)}{4\delta\bar{n}}+2
    w^2\sqrt{\frac{2(1+\xi)}{\delta Z_{\rm eff}}}\right]},
\label{eq_fri}
\end{equation}
where $H$ denotes the Heaviside step function.
Note that we must restrict to $\delta \ll 1$, therefore the high
energy tail of the distribution $f_{RI}$ peaks around $\xi=1$. The
solution is valid for small $w$ or short times $\tau \ll 1$, but only
holds for $w\leq 1$ when $\tau \geq 1$.  The distribution must be
normalized so that $n_i={N}(\tau)\int f_{RI}d^3v$, where the
time-dependent coefficient ${N}(\tau)$ should reduce to the Maxwellian
value as $\tau\rightarrow 0$, ${N}(\tau=0)={N}_0=n_i/(\sqrt{\pi}
v_{Ti})^3$.  By $\tau \sim 1$, the runaway population reaches only a
few percent for $\delta\ll 1$, so the time dependence of the
normalization constant can be neglected and we take ${N}(\tau)\simeq
{N}_0$.

\begin{figure}[htbp]
\begin{center}
  \includegraphics[width=0.49\textwidth]{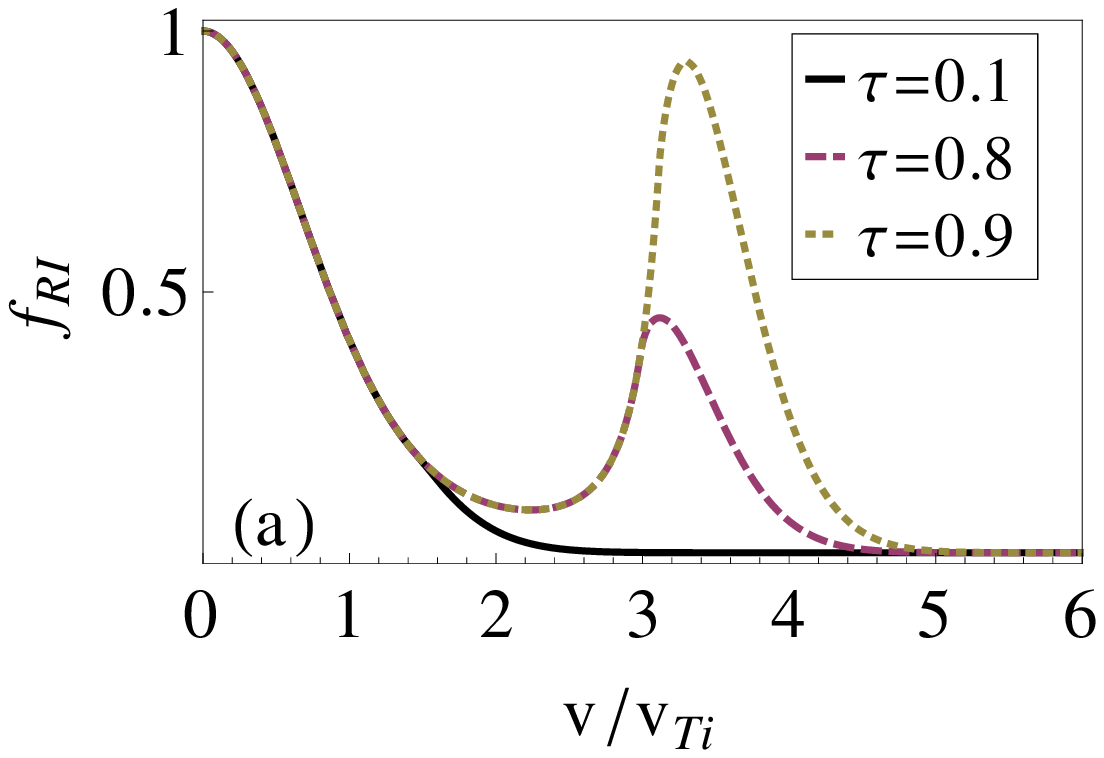} \includegraphics[width=0.49\textwidth]{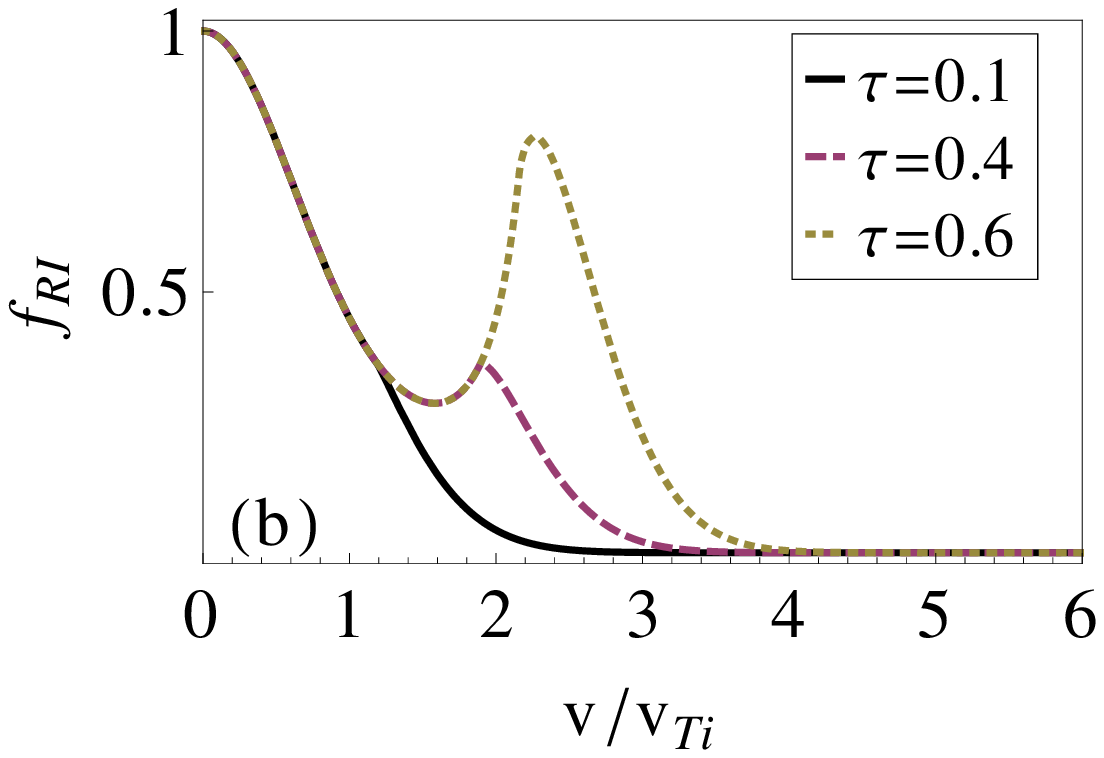} \caption{Runaway
    ion distribution for $\xi=1$ from Eq.~(\ref{eq_fri}) as a function
    of the deuterium speed normalized to the thermal speed for
    $\bar{n}=1$ (a) and $\bar{n}=0.5$ (b). Here $\delta=0.1$ and $T=
    10$ eV. }
      \label{fri}
\end{center}
\end{figure}

Figure \ref{fri} shows the effect of $\bar{n}$ on the evolution of the
runaway ion distribution, for pitch-angle $\xi=1$.  In the trace
impurity case, the runaway piece of the distribution is well separated
from the bulk by the time it reaches significant density.  The
characteristic velocity $v_m$ is much lower at high impurity density
and the large electron density decreases the normalized timescale on
which the distribution is set up.  For typical experimental parameters
$\tau_s$ is a fraction of a microsecond, so the time to establish the
runaway distribution is very short.  {If,} as the runaway ion population
builds up, Alfv\'enic instabilities are excited the analysis
leading to the expression $f_{RI}$ will start to break down.
Therefore, we consider only the initial phase of the wave-particle
interaction and potential instability drive.

{\it Growth rate of Alfv\'en eigenmodes}
The TAE perturbation is typically dominated by two neighbouring
toroidally coupled harmonics at large aspect ratio, with poloidal mode
numbers $m$ and $m+1$ \cite{chengchance}.  The mode is localized about
the minor radius $r=r_0$, at which the magnetic safety factor
$q_0=(2m+1)/2n$ and has a frequency $\omega = v_A/(2 q_0 R_0)$, where
$R_0$ is the radius of the magnetic axis.
The normalized contribution to the linear growth rate of a TAE from a low collisionality population with distribution $f$ is \cite{bettifreidberg,fulopppcf}
\begin{equation}
\frac{\gamma_i}{\omega}=\frac{2 \pi^2 \mu_0 m_i^2 q_0^3 R_0}{B_0^2}\int_0^\infty d v_\perp v_\perp \sum_{v_r=v_A,v_A/3} \frac{v_r}{v_A}\left(v_r^2+\frac{v_\perp^2}{2}\right)^2\left(\omega
\frac{\partial}{\partial \mathcal E}-\frac{n}{q_i}\frac{\partial
}{\partial \psi}\right)f\left|_{|v_\parallel|=v_r}\right. ,
\label{eq_drive}
\end{equation}
where $\mathcal E=m_j v^2/2$, $\psi$ is the poloidal flux and at large
aspect ratio, $d\psi \simeq R B_\theta dr$.  {The two component
  harmonics allow resonant interaction for $|v_\parallel|\simeq v_A/3$
  or $|v_\parallel|\simeq v_A$, so both runaway ions and electrons,
  which have opposite velocities, may drive the mode.}  For brevity,
and motivated by the observations of~\cite{zeng}, we specifically
consider stability conditions for the case where both $n$ and $\omega$
are positive, but note that alternative combinations can be of
interest~\cite{sandquistetal2007}.

\begin{figure}[htbp]
\begin{center}
  \includegraphics[width=0.49\textwidth]{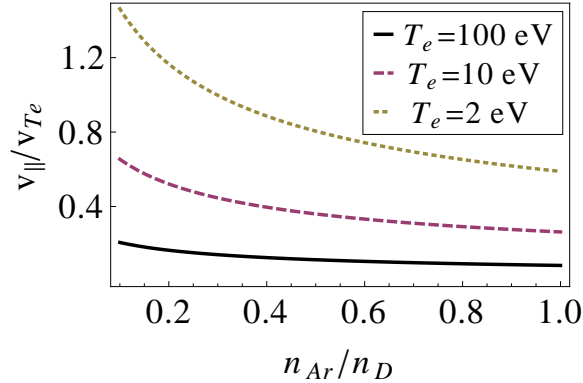} \caption{
    Normalized deuterium ion speed satisfying the $v_{\parallel}=v_A/3$ resonance as a function of the assimilated
    argon content for
    $B=2\;\rm T$, $n_D=2\times 10^{19}\;\rm m^{-3}$ and $n_C=0.08 n_D$.}
       \label{figresonance}
\end{center}
\end{figure}

As the runaway ions accelerate, the inverted region of their energy
distribution will reach the lower Alfv\'{e}n resonance and drive the
TAE via the $\partial f/\partial \mathcal{E}$ term.  In addition, if
the radial runaway ion profile peaks on axis, the term proportional to
$\partial f/\partial \psi$ will give a positive contribution to the
growth rate. 
For $n_{Ar}=0.1 n_D=2\times 10^{18}\;\rm m^{-3}$, $n_{C}=0.08 n_D$, $B=2T$
and $T_e=10\;\rm eV$ the condition $v_\parallel=v_A/3$ requires ions
with velocities $v_{RI}\simeq 0.65 v_{Te}$.
For typical low $m \sim 1$, $q_0 = 1.5$ and $R_0 = 1.75\; \rm m$, ions
with these velocities (or above) would drive a TAE with frequency
$112\;\rm kHz$, which is in the frequency range of the
observations reported in \cite{zeng}.  Note that a higher amount of
assimilated argon or a lower magnetic field would lead to a lower
Alfv\'en velocity and TAE frequency, and in that case the resonance
condition with runaway ions would be more readily met. Figure
\ref{figresonance} shows the normalized deuterium speed that is
required to fulfill the resonance $v_\parallel=v_A/3$ for various
temperatures.

At low temperatures, the analytical form of the distribution given in
Eq.~(\ref{eq_fri}) breaks down before the ions accelerate sufficiently
and the growth rate must be calculated using a numerical solution of
the ion kinetic equation.  However, we give an illustrative analytical
calculation of the runaway drive for the same situation at $T_e =
500\;\rm eV$, with $\delta = 0.1$ and $Z_{Ar} = Z_C = 2$. 
 The electric field builds up as the plasma temperature is quenched in
 the disruption, and so it is also possible that such a resonant
 interaction can begin during the cooling process.  
The gradient with respect to energy is related to that with respect to
$w$ as $\partial f_{RI}/\partial \mathcal
E=(2\delta/m_iv_{Ti}^2w)(\partial f_{RI}/\partial w)$.  Restricting
the runaway density to a few percent allows gradients arising from the
normalization constant to be neglected, and taking $\xi \approx 1$ at
the resonance, only the term in $v_r^5$ contributes in
Eq.~(\ref{eq_drive}).  At $\tau = 2.3$, the runaway density is $3.9\%$
of the initial bulk ion density and the normalized drive at the
$v_A/3$ resonance is $4.1\%$.  {Note that although fast ions
  resulting from fusion reactions or auxiliary heating often excite
  high-frequency instabilities (magnetosonic or ion cyclotron waves),
  ion runaways are not likely to reach high enough velocities to
  become resonant with higher frequencies and excite such waves. }

{\it Electron runaways}
Finally, we consider the runaway electron population.  Typically it
does not have an inverted energy gradient \cite{hollmann} and whilst
anisotropy in pitch-angle is known to drive instability
\cite{fulop2006}, this will not be effective for the low frequency
modes of interest here.  From Eq.~(\ref{eq_drive}) however, we see
that TAE modes can also be driven by resonant interaction with fast
electrons with steep density profiles. Such profiles may arise at
radially localized current sheets, which are often formed due to a
thermal instability resulting from a balance between heat-diffusion,
radiation and Ohmic heating \cite{putv}.  In hot plasmas, typically
$v_A < v_{Te}$ and the bulk electrons produce damping.  However in
cold plasmas the primary resonance condition $v_\parallel = v_A$ moves
out to velocities greater than the electron thermal speed.  Then we
may expect energy transfer to the mode from the well populated lower
energy part of the runaway electron distribution.  We speculate that
kinetic corrections to the TAE mode structure \cite{fasoli} may allow
interaction to occur with the electron population over a range of
energies in this region.  This would give rise to a distribution in
the observed magnetic fluctuation frequencies and so may be of
relevance in understanding the similar observations discussed in
\cite{zeng}. The higher the magnetic field, the higher the Alfv\'en
velocity, so fewer REs will be able to fulfil the resonance. This
reduces the growth rate and may be the explanation for the
disappearance of the fluctuations at high magnetic fields.
{Note that the situation is complicated by the fact that the
  Alfv\'{e}n velocity is heavily influenced by the impurity content.
  Experimental observations at JET show that traces of runaways can be
  found for argon massive gas injection even at magnetic fields down
  to 1.2 T \cite{lehnen}, whilst for neon injection much weaker
  runaway signatures are observed.  This may indicate that the
  Alfv\'{e}n speed remains near the electron thermal speed in the
  presence of the heavier impurity, preventing resonance and allowing
  the runaway beam to form.}
	
If the {electron} runaway generation is dominated by primary generation, the
suprathermal part of the electron distribution can be approximated by
$f_e^{\mathrm{RE}}(w_{e\|},w_{e\perp}) \simeq C_e(\ln{w_{\parallel
    e}})^{-1} \exp\left\{ -w_\perp^2/[2(1+Z_{\rm eff})
  \ln{w_{\parallel e}}] \right\}$, in the nonrelativistic limit
\cite{cohen}. Here $w=v(2E)^{1/2}/v_{Te}$, $E=|E_\parallel|/E_D$,
the normalization constant is $C_e=n_{re} E/(\pi v_{e,max}
v_{Te}^2(1+Z_{\rm eff}))$, $n_{re}$ is the runaway electron
density and $v_{e,max}$ is the maximum parallel velocity (which is
time dependent). Taking only the $v_r=v_A$ resonance, the TAE growth
rate due to the spatial inhomogeneity can be rewritten as
\begin{equation}
\frac{\gamma_e}{\omega}=\frac{v_A}{|\omega_{ce}|}\frac{2 \pi^2
  n q_0^4} {\epsilon_0}\frac{m_e}{m_i}\frac{v_Av_{Te}^2}{n_0}\int
dx_\perp x_\perp
\left(1+\frac{x_\perp^2}{2}\frac{v_{Te}^2}{v_A^2}\right)^2
\frac{\partial f_e}{\partial r}\left|_{v_\parallel=v_A}\right.,\end{equation}
where $\omega_{ce}$ is the electron cyclotron frequency, $\epsilon_0=r_0/R_0$, $x$ is the velocity
normalized to $v_{Te}$ and $n_0 = \rho_m /m_i$. This may be evaluated approximately using $f_e^{RE}$, by assuming that the spatial variation of the runaway distribution dominates any variation in the background plasma, which gives, {with $w_{\parallel A}=v_A(2E)^{1/2}/v_{Te}$,}
\begin{equation}
\frac{\gamma_e}{\omega} [\%] = \frac{v_A}{|\omega_{ce}|}\frac{ \pi
  n q_0^4 } {\epsilon_0 \bar{v}_{e,max}}\frac{m_e}{m_i}\left(1 + \frac{v_{Te}^2}{v_A^2}\frac{(1+Z_{\rm eff})\ln{w_{\parallel A}}}{E} + \frac{v_{Te}^4}{2 v_A^4}\frac{(1+Z_{\rm eff})^2(\ln{w_{\parallel A}})^2}{E^2}\right)\frac{n_{re,17}}{n_{0,19}} \frac{1}{L_n^p},
\end{equation}
where $n_{0,19}$, $n_{re,17}$ are the bulk and fast electron densities
expressed in units of $10^{19} {\rm m}^{-3}$ and $10^{17} {\rm
  m}^{-3}$ respectively, $\bar{v}_{e,max} = v_{e,max}/v_A$ {and
$1/L_n^p=\left[\partial_r n_{re}/n_{re} - \partial_r
  \bar{v}_{e,max}/\bar{v}_{e,max}\right]$.}
{The radial derivative of the maximum velocity is expected to have
  opposite sign to the radial variation of the runaway density, so
  enhances the instability growth rate in the case of destabilization.
  In the event of thermal instability, when radially localized current
  sheets are formed, the runaway electron density gradient is very
  steep and the term corresponding to the variation of the maximum
  energy is negligible in comparison.}
	{If secondary generation dominates, the
  distribution function from Ref.~\cite{fulop2006} can be used and the
  growth rate becomes
\begin{equation}
\frac{\gamma_e}{\omega} [\%] = \frac{v_A}{|\omega_{ce}|}\frac{ \pi
  n q_0^4 v_A} {\epsilon_0c_zc }\frac{m_e}{m_i}e^{-\frac{v_A}{c c_Z}}\left(1 + \frac{2 c }{\alpha v_A}+\frac{2 c^2}{\alpha^2 v_{A}^2}\right)\frac{n_{re,17}}{n_{0,19}} \frac{1}{L_n^s},
\end{equation}
 where $\alpha=(\hat{E}-1)/(1+Z_{\rm eff})$, $\hat{E}=E m_e c^2/T_e$,  $c_z=\sqrt{3(Z_{\rm
    eff}+5)/\pi}\ln{\Lambda}$ and $1/L_n^s=\partial_r n_{re}/n_{re}$.}  The growth rate is expected to be
significant at radial locations with large safety factors $q_0$ and
short spatial gradient scale lengths.

Any drive will be countered by various damping mechanisms.  The
remaining bulk ions and electrons will be collisional at the low
temperatures considered, so we do not expect that they will produce
strong resonant damping.  However, collisional damping cannot be
evaluated from Eq.~(\ref{eq_drive}) and would have to be determined by
a collisional treatment of the bulk species.  {In the case of
  runaway electron driven TAE, the energy gradient is also expected to
  damp the wave, as the runaway electron population usually has a
  monotonically decreasing energy spectrum \cite{hollmann}. Alfv\'enic
  instabilities are then only destabilized if the effect of the
  spatial gradient is dominant. } Also, as the parameter profiles
  evolve during the thermal quench and post-disruption, we may expect
  that continuum damping of the mode \cite{berk} can arise and to
  calculate this would require simulation of the detailed TAE
  structure.  Such damping calculations are beyond the scope of this
  work.

{\it Conclusions} In this paper we have considered general forms of
the distributions of runaway particles, which have the potential to
excite low frequency Alfv\'enic instabilities. With regard to the
observations of spontaneous magnetic fluctuations appearing in fusion
plasma disruptions, we find that a variety of mechanisms allow TAE
modes to be driven unstable by runaways.  A steep spatial gradient of
runaway electrons is effective in low temperature plasmas, such as
those typical of post-disruption conditions in tokamaks, whilst
runaway ions form an inverted energy distribution and can give an
effective drive in impure plasmas at higher temperatures ($\mygtrsim
100 \;\rm eV$). In both cases, the drive is most efficient for low
magnetic fields and can therefore account for the absence of
instability at higher magnetic fields.  The magnetic perturbation
associated with the instability is expected to scatter the runaway
electrons and in certain cases may therefore stop beam
formation. Observation of {the} excited waves can also be used for
diagnostic purposes.

{\em Acknowledgements} The authors are grateful to J Connor, A Stahl,
J Hastie, P Helander, E Hollmann and G Papp for fruitful discussions.

\bibliographystyle{unsrt}

\end{document}